\documentclass[journal, draftcls, 12pt, onecolumn]{IEEEtran}
\ifCLASSINFOpdf
\usepackage[pdftex]{graphicx}
\else
\usepackage[dvips]{graphicx}
\fi
\usepackage{amsfonts}
\usepackage{amsmath}
\usepackage{algorithmic}

\begin{document}
%
\title{Adaptive Switching Between Single/Concurrent Link Scheme in Single Hop MIMO Networks}
%
%
%

\author{Pengkai Zhao, and Babak Daneshrad 
  \thanks{This paper was presented in part at Milcom 2011, Baltimore, November 2011.}
  \thanks{Pengkai Zhao is working as a senior engineer at Qualcomm Inc., Santa Clara, CA. This work was done when he was a graduate student at the Department of Electrical Engineering, UCLA, Los Angeles, CA, USA.}
 \thanks{Prof. Babak Daneshrad is with the Department
  of Electrical Engineering, UCLA, Los Angeles, CA, USA.}}%

\maketitle

\begin{abstract}
Concurrent link communications built on multi-antenna systems have been widely adopted for spatial resource exploitation. MIMA-MAC, a classical MIMO MAC protocol utilizing concurrent link scheme, is able to provide superior link throughput over conventional single link MAC (under certain isolated link topologies). However, when utilizing rich link adaptation functions in MIMO systems, there exists a non-ignorable probability that MIMA-MAC's throughput will be lower than that of single link scheme (such probability is dominated by the statistics of instantaneous link topology and channel response). Inspired by this critical observation, and for adapting to various 
link topologies, this paper will present a novel MAC design that can adaptively switch between single or concurrent link scheme. With the aim of absolutely outperforming the single link MAC, here our optimization criterion is to guarantee a throughput result that is either better than or at least equal to single link MAC's counterpart. To highlight the design rationale, we first present an idealized implementation having network information perfectly known in a non-causal way. Then for realistic applications, we further develop a practical MAC implementation dealing with realistic system impairments (distributed handshaking and imperfect channel estimation). Simulation results validate that link throughput in our MAC is higher than or equal to single link MAC's counterpart with minimized outage probabilities. And for ergodic link throughput, our proposed MAC can outperform the single link MAC and MIMA-MAC by around 20{\%}-30{\%}.
\end{abstract}


%
\IEEEpeerreviewmaketitle

\section{Introduction}
%
%
%
%
Multi-antenna systems have been ubiquitously applied in wireless 
communications for boosting the throughput performance or cancelling the 
co-channel interference. Originally, the initial development of MIMO (Multi-Input Multi-Output) systems usually targets at point-to-point single link systems. However, given that MIMO system 
has the ability of suppressing the co-channel interference, concurrent link 
scheme naturally becomes a feasible solution, where multiple independent and parallel links can 
simultaneously transmit their packets in a concurrent way. For spatial capacity enhancement, concurrent link scheme is being considered as a powerful candidate in various network scenarios (e.g., cellular 
network, Ad Hoc network or mesh network). Unfortunately, such scheme also introduces a set of design challenges for system development, which include physical layer algorithms, medium access control (MAC) mechanisms, or even a joint consideration of these two layers.

Before introducing our proposed design, we first look at existing works in the literature relying on concurrent link scheme. To begin with, it is easy to see that concurrent link scheme is a natural solution for cellular networks (\cite{1_multiuser_SDMA}, \cite{2_SDMA_overview}, \cite{VT_SDMA}), 
where base stations with MIMO functions can use distinct beam patterns to simultaneously support 
multiple UEs (User Equipment). However, parallel links in cellular networks often share a 
common Tx or Rx node (i.e., base station), which can greatly ease the 
management of these links. Conversely, this paper focuses 
on more universal a case, where different links are independently 
located without sharing any node (which is an emerging scenario in modern wireless networks). At the same time, using existing PHY layer techniques, some works in the literature have evaluated the concurrent link scheme by comparing various signal 
processing algorithms and calculating associated Shannon capacities. For 
instance, Chen \textit{et al.} \cite{3_MIMO_AdHoc} evaluates the sum throughput of concurrent 
link scheme by calculating network's asymptotic spectral efficiency 
under different MIMO configurations. Ma \textit{et al.} \cite{4_MMSE_capacity} investigates the concurrent 
link scheme by statistically calculating MMSE detection's post-processing SNR. Unfortunately, one critical 
shortage in these works is that they all lack an explicit MAC design 
for managing the resource of concurrent links, and they also lack a concrete MAC 
policy for regulating the access of these links. Hence, for concurrent link scheme, it is strongly recommended to design the MAC and PHY algorithms in a joint way, which has been introduced in the literature as SPACEMAC (\cite{5_SPACEMAC},\cite{6_MIMOMAN}), NULLHOC (\cite{7_NULLHOC},\cite{8_nullhoc}), Net-Eigen MAC \cite{Neteigen} and MIMA-MAC \cite{9_MIMA_MAC} protocols. Specifically, SPACEMAC, NULLHOC and Net-Eigen MAC all aim at 
designing elaborative beamforming vectors to distinguish separate links, while MIMA-MAC focuses on using spatial 
multiplexing and linear Rx vectors to suppress concurrent links' co-channel interference. Here the former three protocols (SPACEMAC, NullHoc and Net-Eigen MAC) essentially rely on the usage of Tx beamforming techniques, which 
is not universally available in practice. It is known that Tx beamforming techniques often require (i) carefully calibrated hardware 
modules, (ii) channel reciprocity between Tx/Rx nodes, and (iii) time-invariant channel response within the packet. These system requirements significantly increase the complexity of these protocols. Due to such complexity consideration, and for practical deployment purpose, 
throughout this paper we focus on spatial multiplexing MIMO systems (\cite{19_11nstd}, \cite{MIMO_overview}), and use MIMA-MAC as our reference MAC protocol.

The key idea in MIMA-MAC is constantly enabling two concurrent links in the network, and each link keeps using half 
of the total spatial streams. Using isolated or representative 
topologies, it has been verified that MIMA-MAC is able to outperform conventional 
single link MAC in terms of link throughput. However, given that MIMO system has rich link adaptation functions, and when taking into account such link adaptation abilities, an in-depth 
comparison between MIMA-MAC and single link MAC reveals that there exists a 
non-ignorable probability that MIMA-MAC's link throughput will be lower than 
that of single link MAC. In other words, performance benefits introduced by 
MIMA-MAC are heavily dependent on the instantaneous link topology and 
channel responses. And unfortunately, MIMA-MAC has little capability in configuring its concurrent links to adapt to instantaneous network environment. In this sense, MIMA-MAC is not a mature design for practical deployment.

Inspired by above observation, this paper will present a novel MIMO MAC 
design that can adapt to instantaneous link topologies and channel 
responses. Such adaptive MAC is interpreted as an intelligent switching 
between single or concurrent link scheme, and the objective is 
to optimize the sum throughput by simultaneously guaranteeing each link's 
throughput to be no less than single link scheme's counterpart. The 
key method in our MAC is to explore MIMO system's optimization 
space located at multiple concurrent links. And the result is that our proposed MAC absolutely outperforms single link MAC because its link throughput is 
either larger than or at least equal to single link MAC's counterpart (with 
minimized outage probability). Particularly, our design comprises two major 
steps. First, relying on ideal and non-causal network 
information, we present an idealized implementation 
highlighting the underlying rationale. Second, considering applications in 
reality, we further develop a practical MAC implementation 
dealing with various system impairments, including distributive handshaking and imperfect channel estimation. Simulation results validate that 
with minimized outage probabilities (that are dramatically lower than those 
of MIMA-MAC), our design can guarantee each link's throughput to be 
either larger than or at least equal to single link MAC's counterpart. 
And in regards to ergodic link throughput averaged from random 
topologies, our design can outperform single link MAC and MIMA-MAC by 
20{\%}-30{\%}.

The reminder of this paper is organized as follows. We introduce the system 
model in section II, and present link adaptation and throughput derivation 
in section III. Comparison between MIMA-MAC and single link MAC is provided in 
section IV. Idealized implementation for our proposed MAC is given in 
section V, and its practical implementation, including distributed handshaking 
and imperfect channel estimation, is discussed in section VI-VII. Simulation 
results are given in section VIII, and conclusions are drawn in section IX.

\section{System Description}
This paper focuses on a single hop MIMO network, where each 
node is within the transmission range of all others. There are multiple 
independent links communicating in the network, resulting in an interference 
limited environment. Link and node locations are randomly and uniformly 
distributed in a rectangle box of 200m by 200m. Every node is equipped with 
$N_A = 4$ antennas for transmission and reception, and uses MIMO-OFDM (Orthogonal Frequency Division Multiplexing) 
systems with $N_C = 64$ subcarriers. Here system bandwidth is $W = 
20\mbox{MHz}$, and OFDM's guard interval is $\rho _G = 1 / 4$. Tx power per 
node is the same and is denoted by $P_T = 25$dBm. We assume that there are a 
total of $K$ links in the network, labeled as link $L_1 $ to link $L_K $. 
Tx and Rx nodes in link $L_q $ are denoted as $T_q $ and $R_q 
$, respectively. Fast fading channel from Tx node $T_q $ to Rx node $R_q $ 
at the $i$th subcarrier is ${\rm {\bf H}}_{R_q ,T_q } (i)$, and power decay between any two nodes is calculated according to simplified path loss model (equation 2.40 in \cite{15_Goldsmith}) with an exponent of 
3, $d_0 $ = 1m, and wave-length $\lambda $ = 0.125m. Here fading channels among 
nodes (including path loss) are generated using 802.11n Channel Model D 
\cite{17_ChanModel}. These channels are static in one Tx frame, but are
independent among different Tx frames. Background noise power per 
subcarrier is defined as $\sigma _N^2 = - 113\mbox{dBm}$.

Every link in the network adaptively selects its stream number subject to 
a maximum value $N_A $. Tx nodes use spatial multiplexing for MIMO transmission (\cite{19_11nstd}, \cite{MIMO_overview}), and Rx nodes use linear vectors for MIMO detection. Besides, every spatial stream adaptively selects its 
modulation and coding scheme (MCS) (including QAM modulation type and channel coding rate). All 8 usable MCSes in this paper are 
listed in Table \ref{Table_MCS}. We denote the QAM symbol at the $i$th 
subcarrier and $m$th stream of node $T_q $ as $X_{T_q } (i,m)$, which 
has zero mean and unit variance. For MAC efficiency 
consideration, every spatial stream can aggregate multiple payload units to fill the 
transmission duration. Such payload units are named as MAC Data Unit (MDU), 
and each MDU has a fixed size of $N_B = 100$ bytes. Also, all MDUs within the 
same stream have the same MCS type. Finally, wireless channels in the 
network are estimated using channel training symbols engraved within packet preamble, and channel estimation details will 
be given in section VII.

Every simulated point is averaged from 1000 independent trials. 
Here one trial represents one random topology realization, while different trails denote independent realizations. We use ${\rm {\bf A}}(i)$ to 
represent the matrix corresponding to the $i$th subcarrier, and ${\rm {\bf 
A}}(i,j)$ is the $j$th column of matrix ${\rm {\bf A}}(i)$. Besides, $[ \cdot ]^H$ represents 
Hermitian calculation. Finally, a list of symbol notations used in this paper are given in Table \ref{param_def}.

\section{Link Adaptation and Throughput Derivation}

\subsection{PPSNR Derivation}

Link throughput in this paper is derived via Post-Processing SNR (PPSNR) values. Consider two concurrent links 
simultaneously transmitting in the network, which are labelled as link $L_1 $ and link 
$L_2 $ (see illustration in Fig. \ref{topo}). Total stream numbers used by these two links are $M_1 $ and 
$M_2 $, respectively. Here we use $\sqrt {1 / M_1 }$ to scale node $R_1 $'s desired 
channel at the $i$th subcarrier and $m$th stream as $\sqrt {1 / M_1 } 
{{\rm {\bf H}}}_{R_1 ,T_1 } (i,m)$ ($1 \le m \le M_1 )$, and 
coefficient $\sqrt {1 / M_1 } $ is to normalize the transmit power. 
Meanwhile, node $R_1 $'s interference channel at the $i$th subcarrier, which 
is caused by node $T_2 $'s $m$th spatial stream, is scaled as $\sqrt {1 / 
M_2 } {{\rm {\bf H}}}_{R_1 ,T_2 } (i,m)$ ($1 \le m \le M_2 )$. 
Using variables defined in section II, the real PPSNR value at the $i$th 
subcarrier and the $m$th stream of link $L_1 $, denoted as $\Gamma _{R_1 
,T_1 } (i,m,M_1 ,M_2 )$, is derived via an MMSE criterion \cite{18_wirelesscomm_David}:
\begin{align}
\label{eq1}
{\rm {\bf W}}_{R_1 ,T_1 } (i,m) &= \sqrt {1 / M_1 } {\rm {\bf C}}_{R_1 ,T_1 
}^{ - 1} (i,m){\rm {\bf H}}_{R_1 ,T_1 } (i,m), \\
\label{eq2}
{\rm {\bf C}}_{R_1 ,T_1 } (i,m) &= \frac{1}{M_1 }\sum\limits_{l = 1,l \ne 
m}^{M_1 } {{\rm {\bf H}}_{R_1 ,T_1 } (i,l)} \left[ {{\rm {\bf H}}_{R_1 ,T_1 
} (i,l)} \right]^H + \nonumber\\
&\frac{1}{M_2 }\sum\limits_{l = 1}^{M_2 } {{\rm {\bf 
H}}_{R_1 ,T_2 } (i,l)} \left[ {{\rm {\bf H}}_{R_1 ,T_2 } (i,l)} \right]^H + 
\sigma _N^2 {\rm {\bf I}}_{N_A } . \\
\label{eq3}
\Gamma _{R_1 ,T_1 } (i,m,M_1 ,M_2 ) &= \frac{1}{D_{R_1 ,T_1 } (i,m)} \cdot 
\frac{1}{M_1 }\left| {{\rm {\bf W}}_{R_1 ,T_1 }^H (i,m) \cdot {\rm {\bf 
H}}_{R_1 ,T_1 } (i,m)} \right|^2, \\
\label{eq4}
D_{R_1 ,T_1 } (i,m) &= {\rm {\bf W}}_{R_1 ,T_1 }^H (i,m){\rm {\bf C}}_{R_1 
,T_1 } (i,m){\rm {\bf W}}_{R_1 ,T_1 } (i,m).
\end{align}
Here ${\rm {\bf W}}_{R_1 ,T_1 } (i,m)$ is the linear Rx vector 
at node $R_1$, and ${\rm {\bf C}}_{R_1 ,T_1 } (i,m)$ denotes the covariance of 
interference plus background noise. Besides, $D_{R_1 ,T_1 } (i,m)$ is the 
residual noise power in the MMSE solution.

PPSNR values in this section are real ones 
because of perfect channel estimate. Such real values are used by our simulation engine to evaluate system performance. Conversely, practical systems only have imperfect channel estimate, and 
their PPSNR values are usually imperfect.

\subsection{QoS based Throughput Metric}

Having derived PPSNR value $\Gamma _{R_1 ,T_1 } (i,m,M_1 ,M_2 )$, now we 
further define a new metric, namely, \textit{effective PPSNR}, which essentially serves as an AWGN-equivalent SNR metric. Using $\Gamma _{L_1 
,{\rm{dB}}} (i,m,M_1,M_2) = 10\log _{10} \left[ {\Gamma _{R_1 ,T_1}(i,m,M_1,M_2)} \right]$, we calculate the \textit{effective PPSNR} value for the $m$th stream of link $L_1 $ ($1 \le m \le M_1)$ as \cite{14_PERmodel}:
\begin{equation}
\label{eq5}
\Gamma_{L_1 ,{\rm{dB}}}^{\rm{eff}} (m,M_1,M_2) = \frac{1}{N_C}\sum\limits_{i=1}^{N_C}\Gamma_{L_1 ,{\rm{dB}}}(i,m,M_1,M_2) - \alpha\times var {[{\Gamma_{L_1 ,{\rm{dB}}}(i,m,M_1,M_2)}]}
\end{equation}
Here variance $var$ is calculated over all subcarriers of the $m$th 
stream, and parameter $\alpha = 0.125$ is fitted offline \cite{14_PERmodel}.

Payload reception at each stream is evaluated via effective PPSNR $\Gamma_{L_1 
,{\rm{dB}}}^{\rm{eff}} (m,M_1,M_2)$ and a QoS based method. Consider one given MCS at 
one spatial stream, if this stream's effective PPSNR is above the 
minimum value that is required for the desired QoS (i.e., 10{\%} packet 
error rate, see Table \ref{Table_MCS}), then we declare that all transmitted MDUs 
within this stream are successfully received. Otherwise, these MDUs are 
assumed to be lost. A more complete treatment of PPSNR can be found in \cite{14_PERmodel}. 
Obviously, each stream's optimal MCS can be adaptively selected according to $\Gamma_{L_1,{\rm{dB}}}^{\rm{eff}} (m,M_1,M_2)$.

\subsection{Link Adaptation}

Link adaptation in this paper is to select the optimal stream number and each 
stream's optimal MCS. Here we look at selecting each stream's optimal MCS. Consider two concurrent links ($L_1 $ and $L_2 )$ that 
are transmitting $M_1 $ and $M_2 $ streams, respectively. Using 8 different MCSes, and with effective PPSNR 
$\Gamma_{L_1,{\rm{dB}}}^{\rm{eff}} (m,M_1 ,M_2 )$ in 
hand, the optimal MCS for the $m$th stream of node $T_1 
$ is selected to be the highest MCS whose PPSNR threshold (Table \ref{Table_MCS}) is lower than 
$\Gamma_{L_1 ,{\rm{dB}}}^{\rm{eff}} (m,M_1 ,M_2 )$. 
Using such optimal MCS, MDU number aggregated at the $m$th 
stream of link $L_1 $ is denoted as $N_{L_1}(m,M_1 ,M_2 )$. Accordingly, with parameter set $(M_1 ,M_2)$, total MDU number summed from all streams of link $L_1 $  is calculated as: 
\begin{equation}
\label{eq6}
N_{L_1}(M_1, M_2) = \sum\limits_{m=1}^{M_1}N_{L_1}(m, M_1, M_2)
\end{equation}
Since each MDU has the same payload size ($N_B = 100$ 
bytes), every link's throughput can be represented via 
its total MDU number.

\section{Comparison between Single Link MAC and MIMA-MAC}
MIMA-MAC in this paper always uses two concurrent links for simultaneously transmission, and each link constantly uses $N_A / 2 = 2$ spatial streams\footnote{For ease of description, this paper always assumes that there are $N_A$=4 antennas per Tx/Rx node.}. Link adaptation in MIMA-MAC is to let each stream adaptively select its optimal MCS for throughput maximization. In this section we compare MIMA-MAC with conventional single link MAC that allows 
only one single link transmission (in any one-hop area). For fair comparison, link adaptation is 
enabled in single link MAC as well, where stream number and each stream's 
MCS are both adaptively selected for throughput maximization. Additionally, one classical example for single link MAC is the DCF mode in IEEE 
802.11 standard (\cite{19_11nstd}, \cite{20_11std}).

\subsection{Comparison Results}

\begin{figure}
\centering
\includegraphics[width=5in]{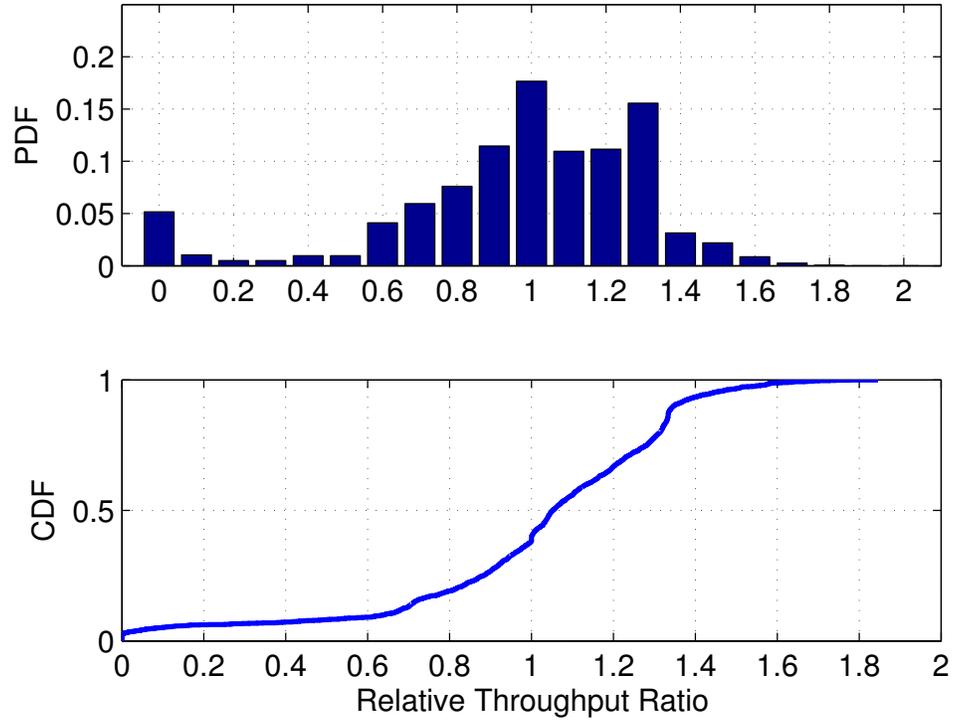}
\caption{PDF and CDF plots for MIMA-MAC's RT ratio values.}
\label{MIMA_res}
\end{figure}

We simulate both single link MAC and MIMA-MAC using the settings 
in section II {\&} III. Specifically, we assume 2 independent 
links in the network (Fig. \ref{topo}). For single link MAC, these 
two links alternatively access the channel in a round-robin manner. And in MIMA-MAC, these two links always access the channel in a concurrent 
manner. We use ideal system conditions for simulations, i.e., MAC layer contention and handshaking overheads are fully ignored, and each link simply uses a time frame with 5ms duration for payload transmission. Also, wireless channels are 
assumed to be perfectly estimated. We investigate MIMA-MAC's relative throughput ratios (RT ratio) at each trial\footnote{Here one trial denotes one random topology realization.} and each 
link. Such RT ratio is defined as the ratio of considered MAC's link throughput compared to that of single link MAC. (This metric represents 
the throughput gain over single link MAC.) We plot in Fig. \ref{MIMA_res} the probability density function (PDF) and 
cumulative density function (CDF) curves of RT ratio values. In PDF plot, the value at the x-axis, say $x_0 $, denotes the 
probability that the RT ratio is distributed within the range of $[x_0 ,x_0 
+ 0.1)$. Results in these figures validate that with a certain probability, 
MIMA-MAC can outperform single link MAC in terms of link throughput, and 
such throughput gain can be as high as 100{\%}. However, under certain topology realizations, MIMA-MAC's 
link throughput is lower than that of single link MAC. The probability 
for such observation is as high as 0.4, which is a non-ignorable value 
for practical applications. Even worse, the lower bound of RT ratio in MIMA-MAC is as poor as less than 0.1.

\subsection{Representative Topologies}

\begin{figure}
\centering
\includegraphics[width=4in]{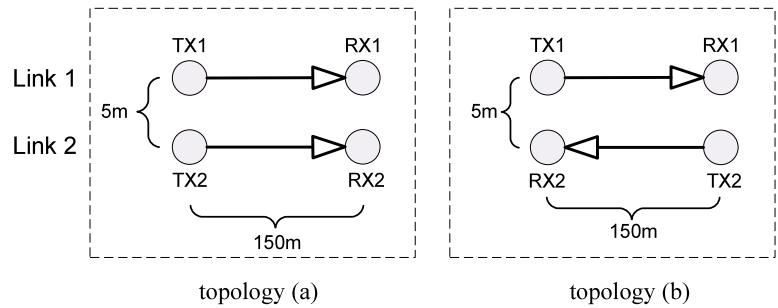}
\caption{Topology examples illustrating our considered network environment.}
\label{topo}
\end{figure}

To further highlight the difference between MIMA-MAC and single link MAC, 
here we look at two motivating topologies (one is for MIMA-MAC's superior performance, and the 
other one is for inherent limitation). The 
first topology is depicted in Fig. \ref{topo}-topology (a), which has two parallel 
links sharing the same transmission direction. Here each 
link's distance is 150m, and the distance between these two links 
is 5m. Using single link MAC, the throughput per link is 17.6 
Mbps. But for MIMA-MAC, its link throughput is 28.4 Mbps, and the throughput gain compared to single link MAC is as high as 61{\%}. The second 
topology is topology (b) in Fig. \ref{topo}, which is similar to the first 
one except that these two links have opposite transmission 
directions. In this case, using single link MAC, each link's throughput is 17.6 Mbps. But for MIMA-MAC and due to serious co-channel 
interference, each link's 
throughput is degraded to be 7.0 Mbps, and MIMA-MAC's 
RT ratio is as low as 40{\%}. Thereby, under certain 
topologies, MIMA-MAC indeed has a lower throughput value compared to single link 
MAC.

\subsection{Design Motivation}

Above discussions have revealed that MIMA-MAC cannot 
always outperform single link MAC in terms of link throughput. For this point, MIMA-MAC is not a mature design because it has 
little capability in using concurrent link scheme to fully outperform 
the single link MAC. In 
this paper, we will present a novel MIMO MAC design that uses instantaneous channel responses to adaptively switch between single or concurrent link scheme. And our objective 
is using concurrent link scheme to provide a throughput performance 
that is better than or at least equal to single link MAC's counterpart. Consequently, in our proposed design, the probability of having lower throughput than single link MAC is minimized to 
be zero (or at least close to be zero).

\section{Idealized Non-causal Implementation}

\subsection{Design Overview}

\begin{figure}
\centering
\includegraphics[width=5in]{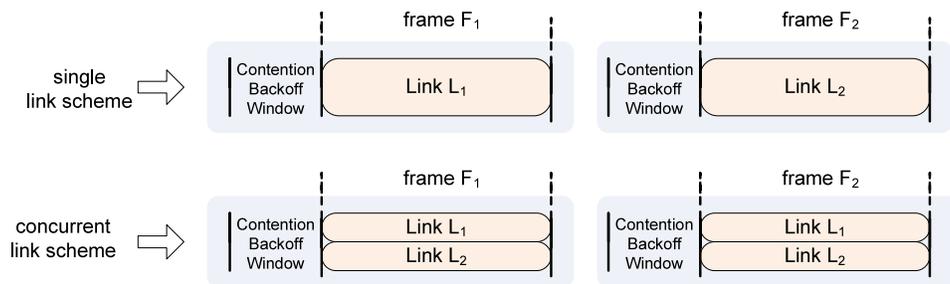}
\caption{Illustration of single link scheme and concurrent link scheme.}
\label{switching}
\end{figure}

This subsection briefly presents the key idea in our proposed design. 
We look at two independent links in the network (link $L_1 $ 
and link $L_2 $), and focus on two separate Tx opportunities (frame $F_1 $ and frame $F_2 $, see Fig. \ref{switching}). For ease of 
description, we name each transmission window as one Tx
frame, which includes handshaking and payload portions, but excludes the contention window\footnote{It should be 
noted that the notion of frame does not necessarily indicate a time division MAC structure.}. Using default single link scheme, we assume that frame $F_1$ is assigned to link $L_1 $, and frame $F_2 $ is for link $L_2 $. We define link $L_1 $'s single link throughput in frame $F_1 $ as $U_{L_1}^S $, and that of link $L_2 $ in frame $F_2 $ is $U_{L_2}^S $. At the same time, concurrent link scheme is defined 
as letting link $L_1 $ and link $L_2 $ simultaneously 
transmit in both frame $F_1 $ and frame $F_2 $ (Fig. \ref{switching}). And under such concurrent link scheme, we use $U_{L_i, F_j}^C $ to denote the throughput of 
link $L_i $ in frame $F_j $ ($1 \le i \le 2$, $1 \le j \le 2)$.

With single/concurrent link schemes in hand, our proposed design is to adaptively switch between these two schemes by satisfying the following optimization criterion: 
\begin{eqnarray}
\label{eq7}
{\rm{Problem (P1): }} \max\ && \sum\limits_{i=1}^{2}\sum\limits_{j = 1}^{2}U_{L_i,F_j}^{C} \\
\mbox{s.t. } \ &&U_{L_1,F_1}^C + U_{L_1,F_2}^C \ge U_{L_1}^S \\
\ &&U_{L_2,F_1}^C + U_{L_2,F_2}^C \ge U_{L_2}^S 
\end{eqnarray}

Such optimization process can be interpreted as using concurrent link scheme 
to improve the sum throughput performance, but at the same time guaranteeing 
each link's throughput to be no less than its single link scheme's 
counterpart. As expected, default single link scheme (with link $L_1 $ 
in frame $F_1 $ and link $L_2 $ in frame $F_2 )$ is a natural candidate 
satisfying conditions (8-9). In this sense, it is safe to expect that 
the solution of problem (P1) will be at least as good as purely using single link scheme.

\subsection{Idealized Non-causal Implementation}
It is important to understand that there is a non-causal assumption in problem (P1). That is, even 
before the start of frame $F_1 $, channel information in both frame $F_1 $ 
and frame $F_2 $ has already become available. This non-causal 
assumption is impractical in reality because it is impossible to 
get frame $F_2 $'s information at the beginning of frame $F_1 $. But here we 
mainly use this assumption to derive performance benchmark. 

For more MAC details, here we bring stream allocations in link $L_1 $ and $L_2 $ into consideration. Assume that in frame $F_1 $, the stream numbers used by link $L_1 $ and $L_2 $ are $M_1^{F_1 } $ and $M_2^{F_1 } $, respectively. Here we 
use superscript $F_1 $ to denote the variables corresponding to frame $F_1 
$, and we use $N_{L_1 }^{F_1 } (M_1^{F_1 } ,M_2^{F_1 } )$ to denote 
the transmission rate of link $L_1 $ in frame $F_1 $, which is the sum of aggregated MDUs at all spatial streams. Other variations, like 
$N_{L_2 }^{F_1 } (M_2^{F_1 } ,M_1^{F_1 } )$, $N_{L_1 }^{F_2 } (M_1^{F_2 } 
,M_2^{F_2 } )$ or $N_{L_2 }^{F_2 } (M_2^{F_2 } ,M_1^{F_2 } )$, can be 
defined in a similar way. Using these notations, link $L_1 $ and $L_2 $'s transmission rates under 
single link MAC, denoted as $N_{L_1 }^{SL} $ and $N_{L_2 }^{SL} $, are given by:
\begin{eqnarray}
\label{eq10}
N_{L_1 }^{SL} = \mathop {\max }\limits_{1 \le M_1^{F_1 } \le N_A } N_{L_1 
}^{F_1 } (M_1^{F_1 } ,M_2^{F_1 } = 0) \\
\label{eq11}
N_{L_2 }^{SL} = \mathop {\max }\limits_{1 \le M_2^{F_2 } \le N_A } N_{L_2 
}^{F_2 } (M_2^{F_2 } ,M_1^{F_2 } = 0)
\end{eqnarray}
Consequently, our proposed adaptive switching, evolving from problem (P1), is 
accordingly defined as:
\begin{eqnarray}
\label{eq12}
{\rm{Problem (P2): }} \max\ && \sum\limits_{i = 1}^{2}\left\{ N_{L_1}^{F_i}(M_1^{F_i}, M_2^{F_i}) + N_{L_2}^{F_i}(M_2^{F_i}, M_1^{F_i}) \right\} \\
{\rm s.t. }\ && \sum\limits_{i=1}^{2} N_{L_1}^{F_i}(M_1^{F_i}, M_2^{F_i}) \geq N_{L_1}^{SL} \\
\ &&\sum\limits_{i=1}^{2} N_{L_2}^{F_i}(M_2^{F_i}, M_1^{F_i}) \geq N_{L_2}^{SL} \\
\ && M_1^{F_i} \geq 0, \ M_2^{F_i} \geq 0 \\
\ && M_1^{F_i} + M_2^{F_i} \leq N_A
\end{eqnarray}

Problem (P2) is a guideline demonstrating our proposed design in a non-causal sense, which is prohibitive from being applied in reality because of its non-causal nature. In the following we will develop a practical and causal implementation that covers distributed handshaking and imperfect channel estimation.

\section{Distributed Handshaking and Practical Implementation}
The key component in practical implementation is a distributed handshaking executed in a causal way. Here we use the 
scenario of two links (link $L_1 $ and link $L_2 )$ and two time frames 
(frame $F_1 $ and frame $F_2 )$ shown in Fig. \ref{switching} to illustrate this handshaking process.

\subsection{Distributive Handshaking in Frame $F_1 $}
\begin{figure}
\centering
\includegraphics[width=5in]{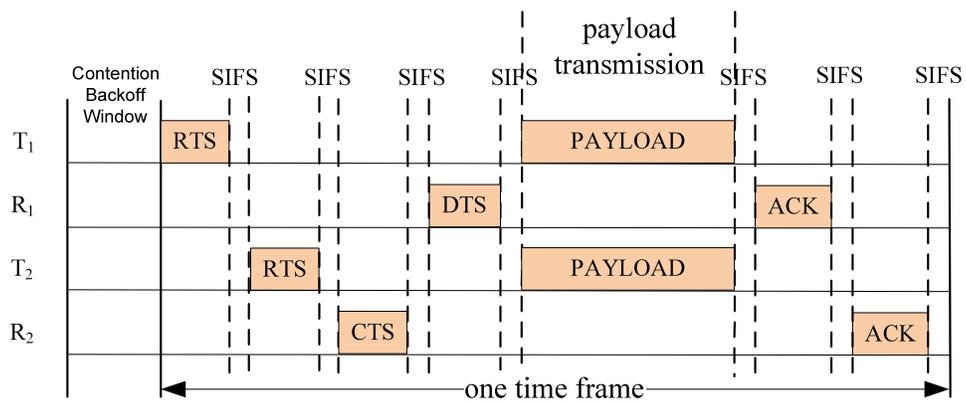}
\caption{Concurrent link handshaking in frame $F_1$.}
\label{con_hand}
\end{figure}

Distributed handshaking in frame $F_1 
$ is depicted in Fig. \ref{con_hand}. After winning the 
contention window, node $T_1 $ and $T_2 $ sequentially send their RTS 
packets for channel learning purpose. Node $R_2 $ learns the channels from 
$T_1 $ (interference channel) and $T_2 $ (desired channel), and estimates 
link $L_2 $'s transmission rates under different configurations (i.e., 
stream numbers used by $T_1 $ and $T_2 $, and each stream's MCS). 
Later, node $R_2 $ uses a CTS packet to inform node $R_1 $ of link $L_2 $'s 
transmission rates under different configurations. At the same time, node $R_1 $ also learns the 
channels from $T_1 $ and $T_2 $, and estimates link $L_1 $'s transmission 
rates under different configurations. Having obtained the feasible rates 
of link $L_1 $ and $L_2 $, node $R_1 $ consequently makes a decision 
between single or concurrent link scheme. Next, node $R_1 
$ broadcasts its switching decision (and MCS configuration) via a DTS 
packet. 

Payload transmission in frame $F_1 $ is determined by switching decision. 
If the decision is concurrent link scheme, then link $L_1 $ and $L_2 $ can 
simultaneously transmit their payload packets (see illustration in Fig. \ref{con_hand}). 
Conversely, if the decision is single link scheme, then in frame $F_1$, link $L_1 $ 
transmits its payload packet via a single link manner, and 
link $L_2 $ refrains from payload transmission.

For causal consideration, since the switching process is executed 
at the beginning of frame $F_1 $, it is impossible to get frame $F_2 $'s 
information at this time point. As a result, switching process in frame $F_1 $ is 
executed via a modified optimization criterion that purely relies on frame $F_1 $'s information. In details, the modified definitions of single link rates 
for link $L_1 $ and $L_2 $ (i.e., $N_{L_1 }^{SL} $ and $N_{L_2 }^{SL} $ in 
section V-B) are calculated as:
\begin{eqnarray}
\label{eq15}
& N_{L_1 }^{SL} = \mathop {\max }\limits_{1 \le M_1^{F_1 } \le N_A } 
N_{L_1 }^{F_1 } (M_1^{F_1 } ,M_2^{F_1 } = 0)\\
\label{eq16}
& N_{L_2 }^{SL} = \mathop {\max }\limits_{1 \le M_2^{F_1 } \le N_A } 
N_{L_2 }^{F_1 } (M_2^{F_1 } ,M_1^{F_1 } = 0)
\end{eqnarray}

The difference between these new definitions (15-16) and the old ones 
(10-11) is that although link $L_2 $'s single link rate, $N_{L_2 }^{SL} $, 
requires frame $F_2 $'s information, here we simply use frame $F_1 $'s 
information to predict its value (\ref{eq16}). Relying on Eqn. (15-16), frame 
$F_1 $'s adaptive switching is executed as: 
\begin{eqnarray}
\label{eq17}
\mbox{Problem (P3): }\max\ && N_{L_1 }^{F_1 } (M_1^{F_1 } ,M_2^{F_1 } ) + N_{L_2 
}^{F_1 } (M_2^{F_1 } ,M_1^{F_1 } ) \\
\label{eq18}
\mbox{s.t. }\ && 2\times N_{L_1 }^{F_1 } (M_1^{F_1 } ,M_2^{F_1 } ) \ge N_{L_1 }^{SL} \\
\label{eq19}
\ && 2\times N_{L_2 }^{F_1 } (M_2^{F_1 } ,M_1^{F_1 } ) \ge N_{L_2 }^{SL} \\
\ && M_1^{F_1} \geq 0, \ M_2^{F_1} \geq 0 \\
\ && M_1^{F_1} + M_2^{F_1} \leq N_A
\end{eqnarray}

Obviously, if there exists an optimal solution with $M_2^{F_1 } > 0$, then 
we should use concurrent link scheme. Otherwise, we simply use single link 
scheme with link $L_1 $ for frame $F_1 $ and link $L_2 $ for frame $F_2 $. 
Note that there is a probability that the above optimization has no 
solution satisfying (18-19). In that case, we simply use the default single 
link scheme.

\subsection{Distributive Handshaking in Frame $F_2 $}
Now we further look 
at frame $F_2 $'s handshaking design, which is fully dependent on the switching decision in 
frame $F_1 $. Naturally, there are two separate possibilities to be 
discussed: (i) frame $F_1 $'s decision is concurrent link scheme; and (ii) 
frame $F_1 $'s decision is single link scheme.

If frame $F_1 $'s decision is single link scheme, then in 
frame $F_2 $, link $L_1 $ should refrain from accessing the channel, and link 
$L_2 $ can transmit its payload packet via a single link manner. In this 
case, frame $F_2 $'s handshaking is essentially a single link 
handshaking allowing only one single link transmission, which is executed via a sequence of RTS, CTS, PAYLOAD and ACK packets (see Fig. \ref{single_hand}). Obviously, such single link handshaking has less MAC overhead compared to concurrent links' counterpart (Fig. \ref{con_hand}).

On the other hand, if frame $F_1 $'s decision is concurrent link scheme, 
then in frame $F_2 $, link $L_1 $ and $L_2 $ have to keep using concurrent 
link handshaking (Fig. \ref{con_hand}). But there is an 
additional consideration for frame $F_2 $'s link configuration. That 
is, intuitively we can directly apply problem (P3) to configure frame $F_2 $'s 
transmission mode, but due to the fact that channels are 
independent in frame $F_1 $ and $F_2 $, there is a possibility that problem (P3), 
solvable in frame $F_1 $, now becomes unsolvable in frame $F_2 $. In other 
words, for frame $F_2 $, single link rates ($N_{L_1 }^{SL} $ and $N_{L_2 
}^{SL} )$ become infeasible to be strictly and simultaneously guaranteed 
(Eqn. (18-19)). As a result, we have to present a new mode 
configuration for frame $F_2 $'s concurrent link 
scheme. 

Here our approach is described as \textit{maximizing the ratio of the 
single link rates that can be guaranteed}. We first define frame $F_2 $'s single link 
rates as: 
\begin{eqnarray}
\label{eq20}
N_{L_1 }^{SL} = \mathop {\max }\limits_{1 \le M_1^{F_2 } \le N_A } 
N_{L_1 }^{F_2 } (M_1^{F_2 } ,M_2^{F_2 } = 0) \\
\label{eq21}
N_{L_2 }^{SL} = \mathop {\max }\limits_{1 \le M_2^{F_2 } \le N_A } 
N_{L_2 }^{F_2 } (M_2^{F_2 } ,M_1^{F_2 } = 0)
\end{eqnarray}

Then we define a new metric, namely, maximum single link ratio, 
$R_{\rm{max}}^{SL}$, which represents the maximum ratio of the single 
link rates that can be guaranteed under frame $F_2 $'s concurrent link 
scheme. This metric is sequentially calculated as follows. First, given stream numbers $(M_1^{F_2 } ,M_2^{F_2 } )$, we use $R_{L_1 }^{SL} (M_1^{F_2 } ,M_2^{F_2 } )$ and $R_{L_2 }^{SL} (M_2^{F_2 } ,M_1^{F_2 } )$ to represent link $L_1$ and $L_2$'s throughput ratios over single link rates:
\begin{eqnarray}
R_{L_1 }^{SL} (M_1^{F_2 } ,M_2^{F_2 } ) = 2\times N_{L_1 }^{F_2 } (M_1^{F_2 },M_2^{F_2 } ) / N_{L_1 }^{SL} \\
R_{L_2 }^{SL} (M_2^{F_2 } ,M_1^{F_2 } ) = 2\times N_{L_2 }^{F_2 } (M_2^{F_2 } ,M_1^{F_2 } ) / N_{L_2 }^{SL} 
\end{eqnarray}
Next, we use $R^{SL}(M_1^{F_2 } ,M_2^{F_2 } )$ to denote the throughput ratio of the two concurrent links, which is the smaller one of $R_{L_1 }^{SL} (M_1^{F_2 } ,M_2^{F_2 } )$ and $R_{L_2 }^{SL} (M_2^{F_2 } ,M_1^{F_2 } )$:
\begin{eqnarray}
R^{SL}(M_1^{F_2 } ,M_2^{F_2 } ) = \min \left\{ {R_{L_1 }^{SL} (M_1^{F_2 } ,M_2^{F_2 } ),R_{L_2 }^{SL} (M_2^{F_2 } ,M_1^{F_2 } )}\right\}.
\end{eqnarray}
Finally, by searching all possible stream allocations ($M_1^{F_2 } \geq 0,  M_2^{F_2 } \geq 0, M_1^{F_2 } + M_2^{F_2 } \leq  N_A $), we get the maximum single link ratio $R_{\rm{max}}^{SL}$:
\begin{eqnarray}
\label{eq22}
R_{\rm{max}}^{SL} = \min\left\{1,{\max\limits_{\scriptstyle M_1^{F_2 } \geq 0,\  M_2^{F_2 } \geq 0 \hfill \atop {\scriptstyle M_1^{F_2 } + M_2^{F_2 } \leq  N_A \hfill}}} R^{SL}(M_1^{F_2} ,M_2^{F_2 } ) \right\}.
\end{eqnarray}
Here we set the upper bound of $R_{\rm{max}}^{SL}$ as 1, meaning guaranteeing at most 100$\%$ of single link rates. In this way, optimization criterion for frame $F_2 $'s concurrent 
link scheme is to maximize the sum throughput by simultaneously maintaining 
the maximum single link ratio $R_{\rm{max}}^{SL} $: 
\begin{align}
\label{eq26}
\mbox{Problem (P4): }\max\ \ \  & \left\{ {N_{L_1 }^{F_2 } (M_1^{F_2 } ,M_2^{F_2 } ) 
+ N_{L_2 }^{F_2 } (M_2^{F_2 } ,M_1^{F_2 } )} \right\} \\
\label{eq27}
{\rm{s.t. }}\ \ \ &R_{L_1 }^{SL} (M_1^{F_2 } ,M_2^{F_2 } ) \ge R_{\rm{max}}^{SL} \\
\label{eq28}
\ \ \ & R_{L_2 }^{SL} (M_2^{F_2 } ,M_1^{F_2 } ) \ge R_{\rm{max}}^{SL} \\
\ \ \ & M_1^{F_2} \geq 0, \ M_2^{F_2} \geq 0 \\
\ \ \ & M_1^{F_2} + M_2^{F_2} \leq N_A
\end{align}

\subsection{Summary}
To summarize our proposed handshaking, initially transmission in 
frame $F_1 $ is executed via a concurrent link handshaking (Fig. \ref{con_hand}), and its adaptive switching is executed via problem (P3) in Eqn. (\ref{eq17}). Given that frame $F_1 $ decides to use concurrent link scheme, 
frame $F_2 $ also uses concurrent link scheme and 
concurrent link handshaking, but it's mode configuration is executed via 
problem (P4) and Eqn. (\ref{eq26}). On the contrary, if frame $F_1 $'s decision is 
single link scheme, then we simply use single link scheme with link 
$L_1 $ in frame $F_1 $ and link $L_2 $ in frame $F_2 $. Finally, an algorithmic diagram illustrating our proposed switching is listed in Fig. \ref{alg_diagram}.

\section{Imperfect Channel Estimation}

\subsection{Channel Estimation Method}

This paper assumes that wireless channels from one Tx antenna to all Rx 
antennas are estimated using $N_T $ training symbols, and different Tx 
antennas' training symbols do not overlap in the time domain. In particular, we use a 
time domain method to estimate the multi-path channel responses in OFDM system (\cite{16_time_chest}, \cite{TD_Kalman}). By adopting this method, and with $N_T $ training 
symbols per Tx antenna, the variance of channel estimation error per 
subcarrier is given by $\frac{L_{\rm max}\sigma _N^2 }{N_C N_T }$, where $L_{\rm max} $ is 
the number of time domain channel paths, and $N_C $ is the number of OFDM 
subcarriers. Note that generally there exists $L_{\rm max} \ll N_C$, hence we have $\frac{L_{\rm max} \sigma _N^2 }{N_C N_T } \ll \frac{\sigma _N^2 }{N_C}$.

\subsection{PPSNR Estimation under Imperfect Channel Information}
This subsection describes the PPSNR values under channel estimation errors. Recall that we use ${\rm {\bf H}}_{R_{k,} T_l }(i) \in {\mathbb C}^{N_A \times 
N_A } (1 \le k \le 2,1 \le l \le 2)$ to denote the wireless channel from node 
$T_l $ to node $R_k $ at the $i$th subcarrier (including path loss). 
Besides, ${\rm {\bf H}}_{R_{k,} T_l }(i,m)$ denotes the $m$th column 
of matrix ${\rm {\bf H}}_{R_k ,T_l } (i)$, representing the channel 
at the $m$th stream of node $T_l $. In practice, such 
channel information is estimated via training symbols in RTS packets (see handshaking in section VI). Given $N_T $ training 
symbols per Tx antenna, the imperfect estimate of channel response ${\rm {\bf H}}_{R_{k,} T_l }(i,m)$ is given by:
\begin{equation}
\label{eq29}
\widehat{{\rm {\bf H}}}_{R_{k,} T_l }(i,m) = {\rm {\bf H}}_{R_{k,} T_l }(i,m) + \sqrt {\frac{L_{\rm max} \sigma _N^2 }{N_C N_T }} {\rm {\bf Z}}_{R_k ,T_l } (i,m).
\end{equation}
Here ${\rm {\bf Z}}_{R_k ,T_l } (i,m)$ is a column vector representing channel estimation error, whose elements are independent white Gaussian variables with zero mean and unit variance.

Now we further consider the calculation of $\widehat{\Gamma }_{R_1 ,T_1 } (i,m,M_1 ,M_2 
)$, which represents the estimated PPSNR value at the $i$th subcarrier and $m$th stream of link $L_1$. Using similar derivation in Sect. III-A, $\widehat{\Gamma }_{R_1 ,T_1 } (i,m,M_1 ,M_2 )$ is calculated as:
\begin{eqnarray}
\label{eq30}
\widehat{{\rm {\bf W}}}_{R_1 ,T_1 } (i,m) &&= \sqrt {1 / M_1 } \widehat{{\rm 
{\bf C}}}_{R_1 ,T_1 }^{ - 1} (i,m)\widehat{{\rm {\bf H}}}_{R_1 ,T_1 } 
(i,m), \\
\label{eq31}
\widehat{{\rm {\bf C}}}_{R_1 ,T_1 } (i,m) &&= \frac{1}{M_1 }\sum\limits_{l = 
1,l \ne m}^{M_1 } {\widehat{{\rm {\bf H}}}_{R_1 ,T_1 } (i,l)} \left[ 
{\widehat{{\rm {\bf H}}}_{R_1 ,T_1 } (i,l)} \right]^H + \nonumber \\
&& \frac{1}{M_2 
}\sum\limits_{l = 1}^{M_2 } {\widehat{{\rm {\bf H}}}_{R_1 ,T_2 } (i,l)} 
\left[ {\widehat{{\rm {\bf H}}}_{R_1 ,T_2 } (i,l)} \right]^H + \sigma _N^2 
{\rm {\bf I}}_{N_A }, \\
\label{eq32}
\widehat{\Gamma }_{R_1 ,T_1 } (i,m,M_1 ,M_2 ) &&= \frac{1}{\widehat{D}_{R_1 
,T_1 } (i,m)} \cdot \frac{1}{M_1 }\left| {\widehat{{\rm {\bf W}}}_{R_1 ,T_1 
}^H (i,m) \cdot \widehat{{\rm {\bf H}}}_{R_1 ,T_1 } (i,m)} \right|^2, \\
\label{eq33}
\widehat{D}_{R_1 ,T_1 } (i,m) &&= \widehat{{\rm {\bf W}}}_{R_1 ,T_1 }^H 
(i,m)\widehat{{\rm {\bf C}}}_{R_1 ,T_1 } (i,m)\widehat{{\rm {\bf W}}}_{R_1 
,T_1 } (i,m).
\end{eqnarray}

\subsection{Impact on Link Adaptation}
Due to channel estimation errors, our derived PPSNR values, $\widehat{\Gamma }_{R_1 ,T_1 } (i,m,M_1 ,M_2 )$, are usually 
different from the real ones in section III. This can impact our link adaptation that heavily relies on estimated PPSNR values. To mitigate such impact, when deriving effective 
PPSNR ${\Gamma }_{L_1 ,{\rm{dB}}} (i,m,M_1,M_2)$, we will add a new parameter (named as SNR 
backoff value) to compensate for the gap between \textit{estimated} and \textit{real} 
PPSNR values. In this way, and given $\widehat{\Gamma }_{R_1 ,T_1 } (i,m,M_1 ,M_2 )$ values, 
the corresponding effective PPSNR is derived as:
\begin{eqnarray}
\label{eq34}
\widehat{\Gamma }_{L_1 ,{\rm{dB}}} (i,m,M_1,M_2) = 10\log _{10} \left[ 
{\widehat{\Gamma }_{R_1 ,T_1 } (i,m,M_1,M_2)} \right] \\
\label{eq35}
{\widehat{\Gamma }}_{L_1 ,{\rm{dB}}}^{\rm{eff}} (m,M_1,M_2) = \frac{1}{N_C }\sum\limits_{i=1}^{N_C}{\widehat{\Gamma }}_{L_1 ,{\rm{dB}}} (i,m,M_1,M_2) \\
- \alpha \times var \left[ {\widehat{\Gamma }}_{L_1 ,{\rm{dB}}} (i,m,M_1,M_2) \right] - {\Gamma_{L_1}^{\rm Backoff}}
\end{eqnarray}
Here $\Gamma _{L_1 }^{\rm{Backoff}} $ is a correction term that makes up 
for the inaccuracy of the PPSNR estimation induced by imperfect channel 
estimation. Its value can be adaptively tuned at run-time using the real MDU error rate calculated via checksum bits.

\section{Simulation Results and Discussions}

\subsection{Simulation Setup}

This section uses numerical results to evaluate various MACs' throughput performance. Given that most of simulation parameters have 
already been discussed in section II {\&} III, here we only introduce some 
additional ones. In details, fast fading channels among nodes are 
generated via 802.11n Channel Model D \cite{17_ChanModel}, and the maximum number of 
time domain channel paths (parameter $L_{\rm max}$ in section VII) is $L_{\rm max}=8$. Also, parameters for contention and handshaking process are 
listed in Table \ref{Table_II_contention}. Simulations are conducted via the 2-link topology illustrated in Fig. \ref{topo}, where link locations are randomly generated at different trials (the 'trail' definition is given in section IV). There are two simulation settings in this section, which are ideal 
system conditions and practical system conditions.

{\rm\bf Ideal System Conditions}. This setting uses idealized and non-causual 
implementation for our proposed design (section V). Here contention and handshaking overheads are ignored, and each link simply uses a 5ms time frame for payload 
transmission. Also, different links are scheduled in a round-robin manner, and channels are perfectly estimated. Results here mainly serve as performance benchmark.

{\rm \bf Practical System Conditions}. This setting uses practical and causal implementation for 
our proposed design, which covers distributed handshaking (section VI) and 
imperfect channel estimation (section VII). Here contention and handshaking 
overheads are fully accounted, and wireless channels are 
imperfectly estimated via training symbols. In particular, the contention process is 
accomplished via IEEE 802.11's CSMA/CA method, and back-off window parameters are given in Table \ref{Table_II_contention}. Besides, each Tx frame's duration (Fig. \ref{switching}) is fixed as 5ms. Results here mainly represent realistic performance 
achievable in practice.

There are three reference MAC protocols in this paper, which 
are single link MAC, Max Sum Throughput (MST) MAC, and MIMA MAC. Link adaptation function is assumed in all these MACs, which adaptively tunes the stream number and each stream's MCS for throughput maximization.

\textbf{Single Link MAC:} This MAC allows only one single link transmission in one Tx frame. 

\textbf{Max Sum Throughput MAC (MST MAC):} MST MAC is similar to our 
proposed MAC except that it has no consideration for guaranteeing single link 
scheme's counterpart. Instead, this MAC simply maximizes the sum throughput 
(i.e., constraints in Eqn. (3-4) are fully ignored). 

\textbf{MIMA MAC:} This MAC has been discussed in section IV. With 4 
antennas per node, here MIMA-MAC always uses 2 concurrent links and 2 spatial streams per link. For simplicity, MIMA-MAC's handshaking efficiency is assumed to be the same with concurrent 
link handshaking's counterpart (Fig. \ref{con_hand}).

Two primary performance metrics in this paper are relative 
throughput ratio (RT ratio) and ergodic link throughput. As aforementioned, RT 
ratio is defined as the ratio of considered MAC's link throughput compared 
to that of single link MAC, which is calculated at every trail and every 
link. Also, ergodic link throughput denotes the mean throughput per trial 
and per link averaged from various topologies.

\subsection{Results under Ideal System Conditions}

\begin{figure}
\centering
\includegraphics[width=5in]{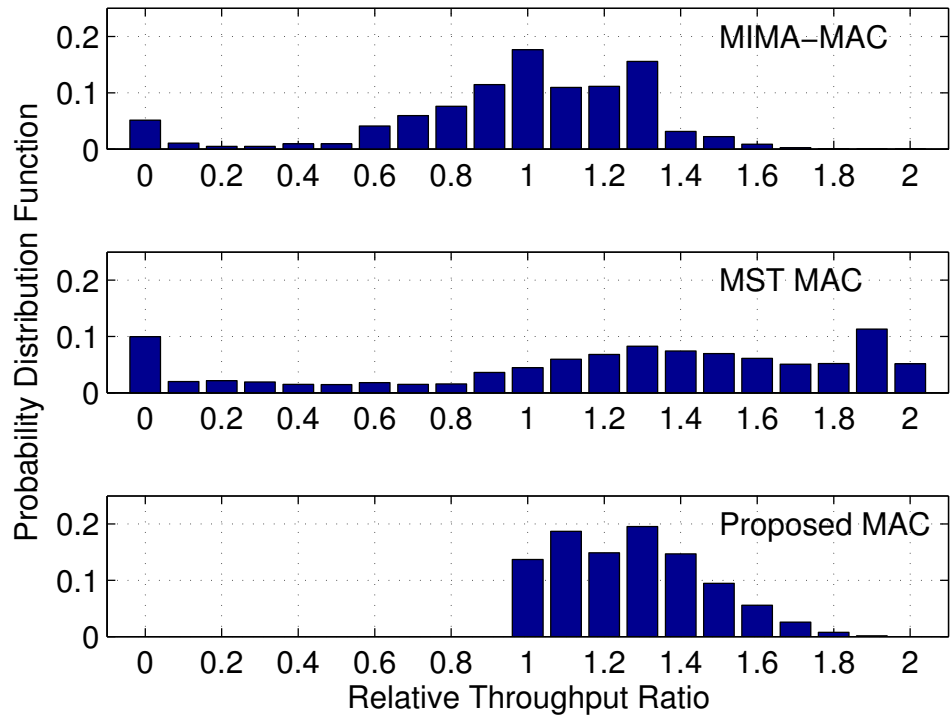}
\caption{RT ratio's PDF values under ideal system conditions.}
\label{ideal_PDF}
\end{figure}

\begin{figure}
\centering
\includegraphics[width=5in]{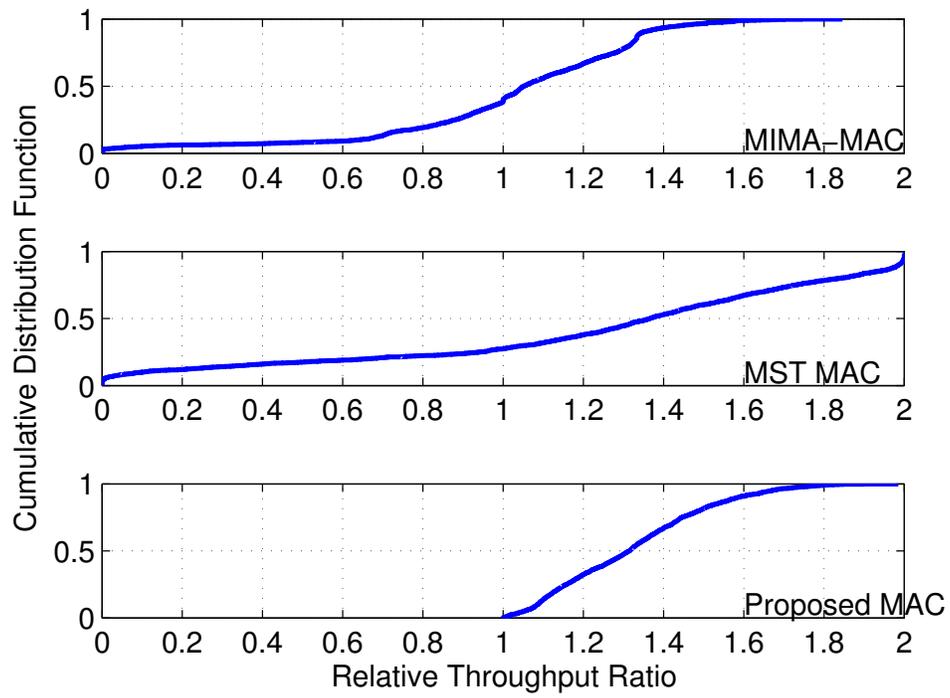}
\caption{RT ratio's CDF curves under ideal system conditions.}
\label{ideal_CDF}
\end{figure}

We start our simulations by investigating throughput performance under 
ideal system conditions. Here we look at 
RT ratio metric and check its PDF (probability distribution function) and CDF (cumulative distribution function) curves, which are collected 
from 1000 independent trials and are plotted in  Fig. \ref{ideal_PDF} and Fig. \ref{ideal_CDF}, respectively. 
Note that for PDF plot, the value corresponding to x-axis label $x_0 $ denotes 
the probability that the associated RT ratio is within the range of $[x_0 
,x_0 + 0.1)$. As expected, using our proposed MAC, the lower bound of 
RT ratio is fixed as 1, indicating that each link's throughput is 
at least no less than single link MAC's counterpart. Besides, the upper bound 
of RT ratio in our MAC is as high as 2. Thereby, our proposed MAC can outperform the single link MAC because its 
performance is better than or at least equal to single link MAC's 
counterpart. But for MIMA-MAC, its RT ratio value is as low as 0. And even worse, there is a remarkable non-zero probability (as high as 0.4) 
that certain link's throughput is lower than that of single link MAC. Thereby, MIMA-MAC has a poor ability in maintaining 
comparable link throughput with single link MAC. Finally, for MST MAC, 
although its RT ratio's upper bound is as high as 2 (meaning an additional 
throughput gain of 100{\%} compared to single link MAC), the associated lower bound is as poor as 0, and the probability of performing worse than single link MAC is as high as 0.3.

After evaluating RT ratio values, now we further look at ergodic link throughput in different MAC protocols (Table \ref{ideal_rate}). Obviously, when compared with single link MAC, our proposed MAC can 
provide an additional throughput gain of 33.6{\%} with respect to ergodic 
link throughput. On the other hand, MIMA-MAC only provides an additional 
gain of around 10{\%} in ergodic throughput over single link MAC. 
This is mostly due to the inefficiency of link adaptation in MIMA-MAC, which 
prohibits it from adapting to various topologies and 
instantaneous channels. This is also why our proposed MAC can 
outperform MIMA-MAC by 23{\%} in ergodic link throughput. Finally, although 
MST MAC has the highest ergodic link throughput, its performance gain comes 
at the expense of degrading certain link's throughput to be as low 
as 0.

\subsection{Impact of System Impairments}
\begin{figure}
\centering
\includegraphics[width=5in]{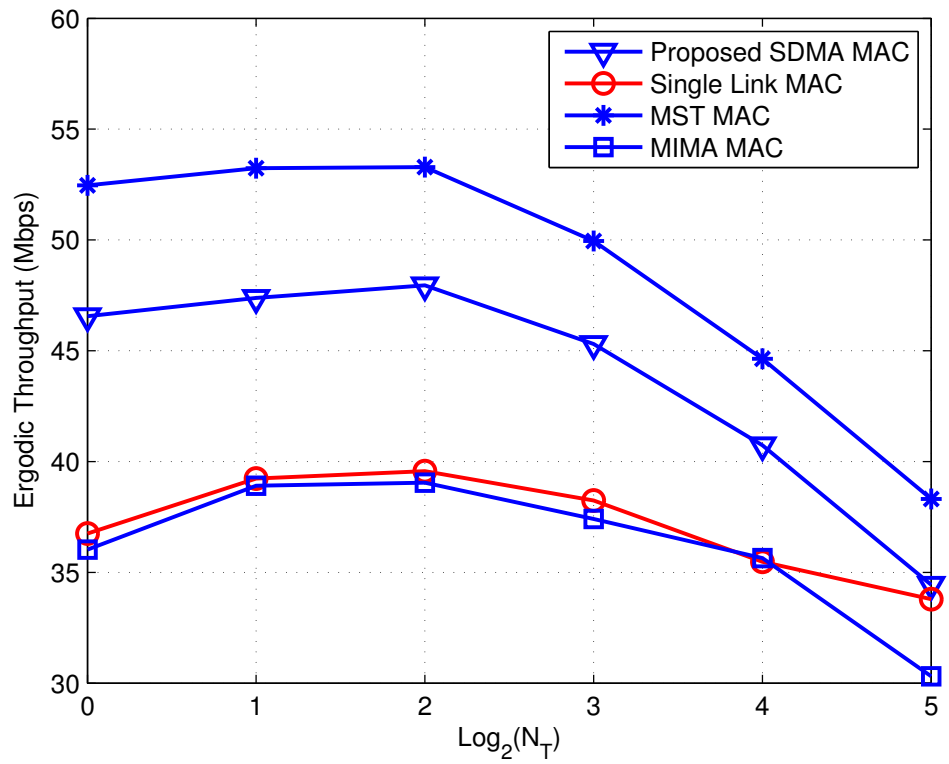}
\caption{Ergodic link throughput versus different training symbol numbers.}
\label{Nt_num}
\end{figure}

This subsection discusses two critical system impairments affecting our proposed MAC, 
which are handshaking overhead and imperfect channel estimation. Since these impairments are closely coupled, here we jointly enable them in the simulation (which include contention overhead, handshaking overhead, and imperfect channel estimation). 
For imperfect channel estimation (section VII), we use ergodic link throughput to select the optimal training 
number $N_T $. It is known that the optimal $N_T $ value is balanced by 
two factors. (i) The larger the number of training symbols, 
the better the channel estimation accuracy. (ii) Overused channel training 
symbols can increase the handshaking overhead and decrease the link 
throughput. Assuming up to 32 training symbols, we plot the ergodic link 
throughput versus different $N_T $ values in Fig. \ref{Nt_num}. It is shown that 
performance summit in all these MAC protocols occurs at 4 
training symbols. Thereby, in this paper we use optimal $N_T $ 
value as $N_T = 4$.

\begin{figure}
\centering
\includegraphics[width=3.5in]{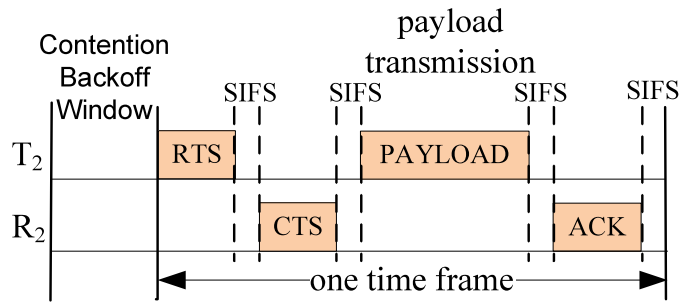}
\caption{Link $L_2$'s single link handshaking in frame $F_2$.}
\label{single_hand}
\end{figure}

In the sequel, we characterize the handshaking efficiency in 
different MAC protocols. Such handshaking efficiency is defined as the 
ratio of payload portion compared to the whole Tx frame duration (Fig. 
\ref{con_hand} {\&} \ref{single_hand}). Recall that there are two separate handshaking schemes in 
section VI, which are single link handshaking (Fig. \ref{single_hand}) and concurrent link 
handshaking (Fig. \ref{con_hand}). Here we list the efficiency 
values in these two handshaking schemes in Table \ref{handshaking}. These values show that 
compared with the efficiency in single link handshaking (95.7{\%}), concurrent 
link handshaking has a lower value of 91.7{\%} because of its increased control packet number.

\subsection{Results under Practical System Conditions}
\begin{figure}
\centering
\includegraphics[width=5in]{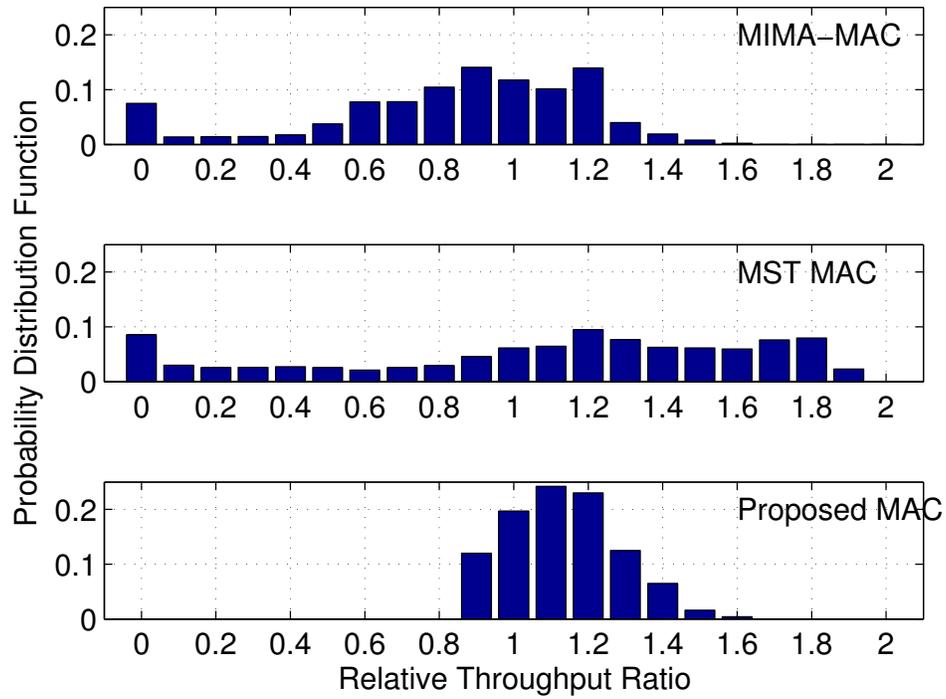}
\caption{RT ratio's PDF values under practical system conditions.}
\label{prac_PDF}
\end{figure}

\begin{figure}
\centering
\includegraphics[width=5in]{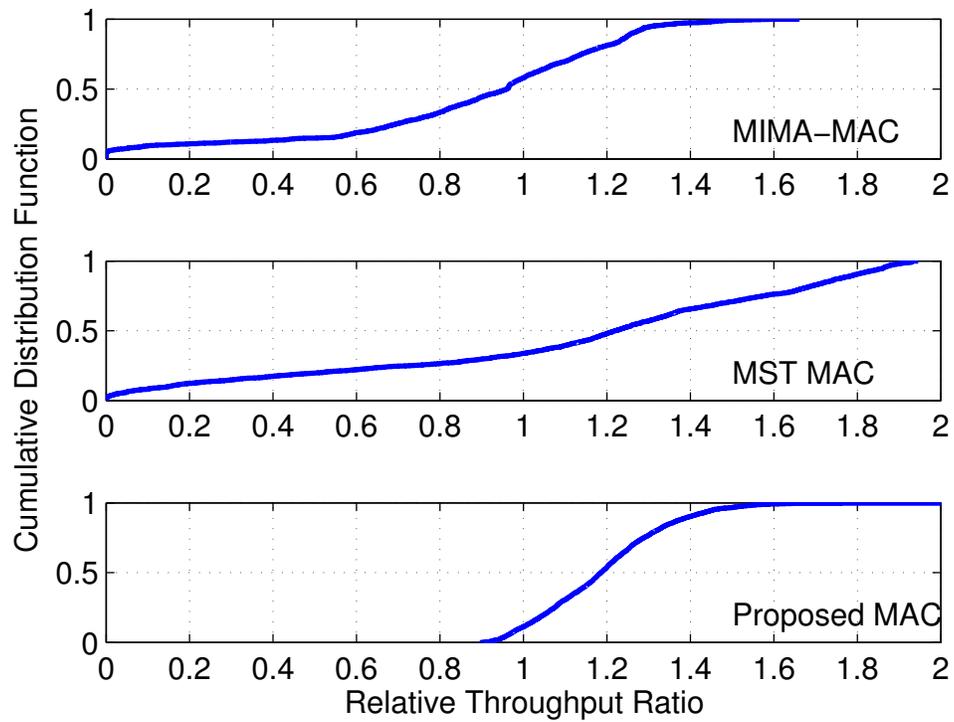}
\caption{RT ratio's CDF curves under practical system conditions.}
\label{prac_CDF}
\end{figure}

Having investigated various system impairments, now we are ready to 
use practical system conditions to evaluate the link throughput in different MAC protocols. Here practical system conditions 
include imperfect channel estimation, MAC 
handshaking overhead, and MAC contention overhead. Before looking at numerical results, we first 
point out that due to system impairments (channel estimation errors and handshaking overhead), there exists a non-zero probability that our proposed MAC's link 
throughput will be lower than that of single link MAC. With this 
point in mind, we evaluate the RT ratio metric by calculating its 
outage probability. We plot RT ratio's PDF and CDF curves collected from 
1000 random trials in Fig. \ref{prac_PDF} and Fig. \ref{prac_CDF}, respectively. Again, in Fig. \ref{prac_PDF}'s 
PDF plot, the value corresponding to x-axis label $x_0 $ denotes the 
probability that the associated RT ratio is within the range of $[x_0 ,x_0 + 
0.1)$. Results in these figures clearly verify that, using practical system 
conditions, and in terms of guaranteeing 100{\%} and 95{\%} of single link 
MAC's throughput, the outage probabilities in our proposed MAC are as 
low as 0.12 and 0.02, respectively. In other words, 98{\%} of the time, our 
proposed MAC's link throughput is larger than 95{\%} of single link MAC's 
counterpart; and 88{\%} of the time, our proposed MAC's throughput is larger 
than that of single link MAC. Conversely, using MIMA-MAC, outage 
probabilities for guaranteeing 100{\%} and 95{\%} of single link MAC's 
throughput are as high as 0.58 and 0.48, respectively. Even worse, 
the lowest RT ratio in MIMA-MAC is close to 0. Thus, under certain topologies, MIMA-MAC indeed
performs worse than single link MAC. Finally, for various topologies, MST MAC also has a poor performance in maintaining better or 
comparable throughput with single link MAC.

At this point, we further investigate the ergodic link throughput under 
practical system conditions, which are averaged from 1000 trials and are 
depicted in Table \ref{prac_rate}. Compared with single link MAC or MIMA 
MAC, and even under practical system impairments, our proposed MAC still 
provides around 20{\%} higher throughput gain in terms of ergodic link throughput. On the contrary, MIMA-MAC's result is close to that of single link MAC, which is caused by its lower handshaking 
efficiency compared to single link MAC. Finally, MST MAC in Table \ref{prac_rate} has the highest ergodic link throughput, but it has no consideration for guaranteeing comparable throughput with single link MAC.

\section{Conclusions}

This paper has presented a novel MIMO MAC design that can adaptively switch 
between single or concurrent link scheme. Here our design objective 
is to absolutely outperform the single link MAC by guaranteeing each link's throughput to be better than or at 
least equal to single link scheme's counterpart. Such adaptive switching is accomplished by 
exploring MIMO system's rich optimization space located at independent and 
concurrent links. And our optimization is built on instantaneous topology information and 
channel response. Using ideal system conditions and non-causal information, 
we first present an idealized implementation illustrating the underlying design rationale. Then for realistic system conditions, we 
further develop a practical and casual implementation covering distributed 
handshaking and imperfect channel estimation. Simulation results verify that 
with minimized outage probabilities (that are significantly lower than those of MIMA 
MAC), our design's link throughput is larger than or at least equal to 
single link MAC's counterpart. And in terms of ergodic link throughput, our 
design can outperform single link MAC and MIMA-MAC by around 20{\%}-30{\%}.


%

\ifCLASSOPTIONcaptionsoff
  \newpage
\fi



%
\bibliographystyle{IEEEtran}
\bibliography{IEEEabrv,tvt_draft}

%








\newpage
\newcommand{\tabincell}[2]{\begin{tabular}{@{}#1@{}}#2\end{tabular}}
\begin{table}
\caption{List of Modulation and Coding Schemes \cite{MCS}} \centering \label{Table_MCS}
\begin{tabular}{|c||c||c||c|}
\hline \textbf{\tabincell{c}{MCS \\ Index}} & \textbf{QAM Type} & \textbf{\tabincell{c}{Coding\\ Rate}} & \textbf{\tabincell{c}{Minimum required effective \\ PPSNR to achieve target \\ BER/PER (10{\%} PER)}} \\
\hline 0 & \textrm{BPSK} & 1/2 & \textrm{1.4 dB} \\
\hline 1 & \textrm{QPSK} & 1/2 & \textrm{4.4 dB} \\
\hline 2 & \textrm{QPSK} & 3/4 & \textrm{6.5 dB} \\
\hline 3 & \textrm{16QAM} & 1/2 & \textrm{8.6 dB} \\
\hline 4 & \textrm{16QAM} & 3/4 & \textrm{12 dB} \\
\hline 5 & \textrm{64QAM} & 2/3 & \textrm{15.8 dB} \\
\hline 6 & \textrm{64QAM} & 3/4 & \textrm{17.2 dB} \\
\hline 7 & \textrm{64QAM} & 5/6 & \textrm{18.8 dB} \\
\hline
\end{tabular}
\end{table}

\begin{table}
\caption{List of Parameter Definitions} \centering \label{param_def}
\begin{tabular}{|c||c|}
\hline \textbf{Parameter} & \textbf{Definition} \\
\hline $N_A$ & antenna number \\
\hline $N_C$ & subcarrier number \\
\hline $W$ & system bandwidth \\
\hline $\rho_G$ & OFDM guard interval \\
\hline $P_T$ & Tx Power \\
\hline $i$ & subcarrier index \\
\hline ${\rm\bf H}_{R_q,T_q}(i)$ & channel response at the $i$th subcarrier \\
\hline $N_B$ & number of bytes in one MDU \\
\hline $M_q$ & stream number in link $q$ \\
\hline $\Gamma$ & PPSNR value\\
\hline $N_{L_1}$, $N_{L_2}$ & aggregated MDU numbers in link $L_1$ and $L_2$ \\
\hline $F_1, F_2$ & frame 1 and frame 2 \\
\hline $N_T$ & training symbol number \\
\hline $L_{\rm max}$ & maximum number of time domain channel paths \\
\hline $\sigma^2$ & background noise power \\
\hline $m$ & spatial stream index\\
\hline
\end{tabular}
\end{table}

\begin{figure}
\centering
\begin{algorithmic}[1]
\STATE {Choose between ideal or practical implementation.}
\IF {this is for ideal and non-casual implemenation}
\STATE {Use ideal system conditions and optimization problem (P2).}
\ELSE
\STATE \COMMENT{Comment: This is for practical and casual implementation.}
\STATE {Use practical system conditions, and frame $F_1$'s adaptive switching is executed via problem (P3).}
\IF {frame $F_1$'s decision is single link scheme}
\STATE {Frame $F_2$ also uses single link scheme.}
\ELSIF {frame $F_1$'s decision is concurrent link scheme}
\STATE {Frame $F_2$ uses concurrent link scheme and optimization problem (P4).}
\ENDIF
\ENDIF
\end{algorithmic}
\caption{The diagram illustrating our proposed adaptive switching process.}
\label{alg_diagram}
\end{figure}

\begin{table}
\caption{Parameters in Contention and Handshaking Process} \centering \label{Table_II_contention}
\begin{tabular}{|c||c|}
\hline \textbf{Handshaking Control Packet} & \textbf{Time Duration} \\
\hline RTS & (6+$N_T\times N_A$)$\times$ 4us \\
\hline CTS & (6 + $N_T$)$\times$ 4us \\
\hline DTS & (4 + $N_T$)$\times$ 4us \\
\hline ACK & (6 + 2)$\times$ 4us \\
\hline \textbf{Contention Parameter} & \textbf{Time Duration} \\
\hline Backoff Time Slot & 9us \\
\hline CWmin & 7 \\
\hline CWmax & 63 \\
\hline
\end{tabular}
\end{table}

\begin{table}
\caption{Handshaking Efficiency Results} \centering \label{handshaking}
\begin{tabular}{|c||c|c|}
\hline {} & \textbf{Single Link Handshaking} & \textbf{Concurrent Link Handshaking} \\
\hline Handshaking Efficiency & 95.7$\%$ & 91.7$\%$ \\
\hline
\end{tabular}
\end{table}

\begin{table}
\caption{Ergodic link throughput under ideal system conditions} \centering \label{ideal_rate}
\begin{tabular}{|c||c|c|c|c|}
\hline {} & \textbf{Proposed MAC} & \textbf{Single Link MAC} & \textbf{MIMA-MAC} & \textbf{MST MAC} \\
\hline Ergodic Throughput (Mbps) & 56.16 & 42.05 & 46.05 & 62.52 \\
\hline
\end{tabular}
\end{table}

\begin{table}
\caption{Ergodic link throughput under practical system conditions} \centering \label{prac_rate}
\begin{tabular}{|c||c|c|c|c|}
\hline {} & \textbf{Proposed MAC} & \textbf{Single Link MAC} & \textbf{MIMA-MAC} & \textbf{MST MAC} \\
\hline Ergodic Throughput (Mbps) & 47.95 & 39.57 & 39.05 & 53.28 \\
\hline
\end{tabular}
\end{table}

\end{document}